\documentclass[12pt,preprint]{aastex}

\newcommand{\kms}{\mbox{km\,s$^{-1}$}}

\slugcomment{To appear in AJ... sometime}

\shorttitle{Chemical Homogeneity in the Hyades}
\shortauthors{De Silva et al.}

\begin{document}

\title{Chemical Homogeneity in the Hyades}


\author{G.M. De Silva\altaffilmark{1}, C. Sneden\altaffilmark{2}, D.B. Paulson\altaffilmark{2},
M. Asplund\altaffilmark{1}, J. Bland-Hawthorn\altaffilmark{3}, \\M.S. Bessell\altaffilmark{1},
K.C. Freeman\altaffilmark{1}}

\altaffiltext{1}{Mount Stromlo Observatory, Australian National University,
Weston ACT 2611, Australia; email: gayandhi@\\mso.anu.edu.au}
\altaffiltext{2}{Dept of Astronomy, University of Texas, Austin, TX 78712;
chris@verdi.as.utexas.edu} 
\altaffiltext{3}{Anglo-Australian Observatory, Eastwood NSW 2122, Australia;
jbh@aao.gov.au} 


\begin{abstract} We present an abundance analysis of the heavy
  elements Zr, Ba, La,
  Ce, and Nd for Hyades cluster F-K dwarfs based on high resolution,
high S/N ratio spectra from Keck/HIRES. The derived abundances show the
stellar members to be highly uniform, although some elements show a small
residual trend with temperature. The rms scatter for each element for the cluster members is as follows;
Zr = 0.055 dex, Ba = 0.049 dex, Ce = 0.025 dex, La = 0.025 dex, Nd = 0.032 dex. This is consistent 
with the measurement errors, and implies that there is little or no intrinsic
scatter among the Hyades members.  Several stars thought to be non-members of
the cluster based on their kinematics are found to deviate from the cluster
mean abundances by about $2\sigma$. Establishing chemical homogeneity in open clusters is
the primary requirement for the viability of chemically tagging Galactic disk stars to
common formation sites, in order to unravel the dissipative history of early
disk formation. 
\end{abstract}





\section{INTRODUCTION}

One of the major goals of near-field cosmology is to
determine the sequence of events involved in the formation of the Galactic
disk during the epoch of dissipation. Since the disk formed dissipatively and
evolved dynamically, most of the dynamical information is lost. However,
locked away within the stars, the
chemical information survives. If star-forming aggregates have unique 
\emph{chemical signatures}, we can use the method of chemical tagging
to track individual stars back to a common formation site \citep{fbh02}. With
sufficiently detailed 
abundances we would be able to reconstruct the stellar aggregates which
have long since diffused into the Galaxy background. The critical issue
for the viability of this method is whether the star-forming aggregates
do indeed have unique chemical signatures. The essential first step is
to investigate the chemical homogeneity of open clusters which are the
likely left overs of star-forming aggregates in the Galactic disk.

\smallskip

Chemical homogeneity in an open cluster implies that the progenitor
cloud was uniformly mixed before its stars
formed. Theoretical work on star formation in giant molecular clouds suggest the
presence of high levels of turbulence \citep{mckeetan02}, which would result
in a well mixed gas cloud. However it is not clear whether this mixing occurs
during cloud collapse (before the birth of the first stars) or wether few
high mass stars form first, shortly after the cloud assembles, and then enrich
the cloud uniformly. With both scenarios however, one expects to find
homogeneity among the long-lived stars in certain key elements.

\smallskip

The heavier neutron-capture elements are the key to establishing chemical
homogeneity, as their abundances are not thought to be modified during normal stellar
evolution. The s-process elements are believed to arise from the He burning
phase of AGB stars, while the most likely site for r-process elements appears
to be type II supernova \citep{b2fh57,wallerstein97}. As a result these
heavier elements (along with some $\alpha$ and Fe-peak elements), are more
representative of the conditions of the progenitor cloud from which the
cluster formed. Further, the neutron-capture elements are
particularly useful for identifying chemical signatures due to the large
intrinsic (cosmic) scatter in their abundances \citep{ed93,reddy03}, in particular at low [Fe/H]. 

\smallskip

Published observations of both light and heavy element abundances in open
clusters demonstrate chemical homogeneity, albeit for only a few stars, and
lend support to the prospect of chemical tagging. Some recent examples include \citet{friel03}, who studied the $\alpha$ elements for four
giants in the old open cluster Collinder 261 and find 1$\sigma$ dispersions
about the mean, consistent with the expected uncertainties. \citet{schuler03}
obtained chemical homogeneity for several elements over nine stars in M34,
except for K which has a tightly correlated
temperature trend, thought to be due to systematic effects in the
model atmospheres. \citet{gonzalez2000} found homogeneity in Eu
for four stars in M11, although there was significant scatter in the $\alpha$
elements. \citet{tau00} studied 9 stars in M67 and found the heavy element
abundances to be almost identical in all stars.
\citet{castro99} observed that the [Ba/Fe] ratios in the Ursa Major
moving group stand out in comparison with solar-neighborhood normal
stars with the same metallicity. 

\smallskip

We begin our study
on chemical homogeneity with the Hyades. It is the nearest cluster to the Sun
with age $\approx$ 625Myrs \citep{Perryman98} and [Fe/H] $\approx$ 0.13
\citep{Pa03}. It
has been extensively studied in the past with stellar memberships firmly
established \citep{debruijne01,hoogerwerfaguilar99}. The Hyades stars have
been subject to numerous abundance studies over past 
years, from \citet{conti65} to most recently \citet{Pa03} (hereafter Pa03).
However to the present date there has been little work
done on the abundance analysis of the heavier neutron-capture elements for the
Hyades dwarfs. Therefore the present results for heavy element abundances over a
large sample of Hyades dwarfs are the first of its kind. 



\section{CHEMICAL ABUNDANCES}
\subsection{Observational Data}
This study on the neutron-capture elements uses a subset of the data sample analyzed
by Pa03. The original observations were part of the planet search program
undertaken with the Keck I/HIRES from 1996 to 2000. A full description of the
observations can be found in \citet{cochran02}.  Those
stars with higher rotation ($v$~sin~$i > 10\kms$) were left out of this study
due to the poorer spectral resolution and blended lines making it difficult to
identify the weak lines of the heavier elements of interest.  

\smallskip

The S/N ratio of each spectrum is typically 100 - 200 per pixel and resolving
power nominally at 60,000. The spectra cover the wavelength region from
3800\AA\ to 6200 \AA, which is an ideal region for the majority of the
neutron-capture lines \citep{bhf04PASA}. Details of the preliminary data
reduction procedures can be found in Pa03.

\subsection{Model Atmospheres and Spectral Lines}
Interpolated Kurucz model atmospheres based on the
ATLAS9 code \citep{Castelli97} with no convective
overshoot were used for this study.

\smallskip

Spectral lines were selected in comparison to the solar atlas
\citep{beckers76}. Preference was given to clean weaker lines covering a
range of excitation potentials. All lines that are formed on or near the
wings of other lines were discarded as well as any lines with significant
blending. Most of the lines used in this analysis have some degree of
blending due to the crowded spectral region where most of the neutron-capture
element lines are formed. The full list of lines and references used for this
analysis is given in Table \ref{lines}. 

\subsection{Abundance Analysis}
Details of the stellar parameter analysis are described in Pa03. The same
parameters have been used for this study as we are using the same data set.  

\smallskip

\begin{figure}
\begin{center}
\includegraphics[scale=0.9, angle=0]{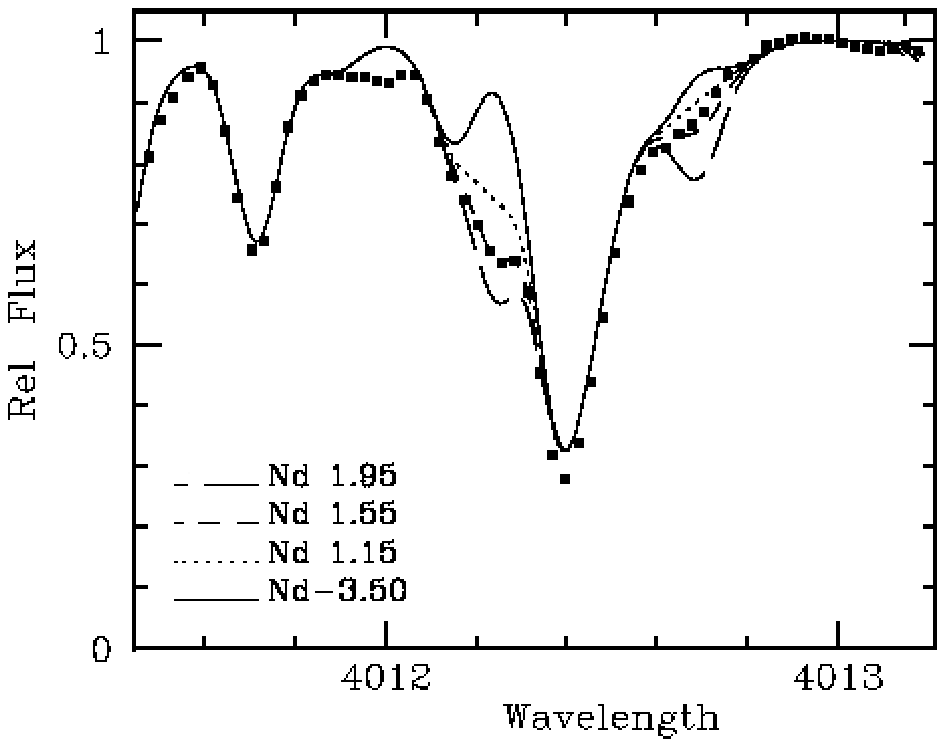}
\caption{Sample synthetic (dashed and dotted curves) and observed (dots)
spectra of the two NdII lines at 4012.24 \AA\ and 4012.69 \AA .}\label{ndsynth}
\end{center}
\end{figure}

The abundance analysis makes use of the MOOG code \citep{sneden73} for LTE
Equivalent Width (EW) analysis and spectral syntheses. Depending on the degree
of blending of the spectral lines, abundances were derived either by EW
measurements or by spectral synthesis. The EWs were
measured by fitting a Gaussian profile to the observed lines of
interest using the interactive $SPLOT$ function in the IRAF\footnote{IRAF is distributed by the National Optical Astronomy
Observatory, which is operated by the Association of Universities for Research
in Astronomy, Inc., under cooperative agreement with the National Science Foundation.} package.
All Zr and Ba abundances were
obtained by EW analysis, while the La, Ce, and Nd lines required syntheses. A sample
synthesis of Nd lines is shown in Figure \ref{ndsynth}.

\smallskip

\begin{figure}[h]
\begin{center}
\includegraphics[scale=0.7, angle=0]{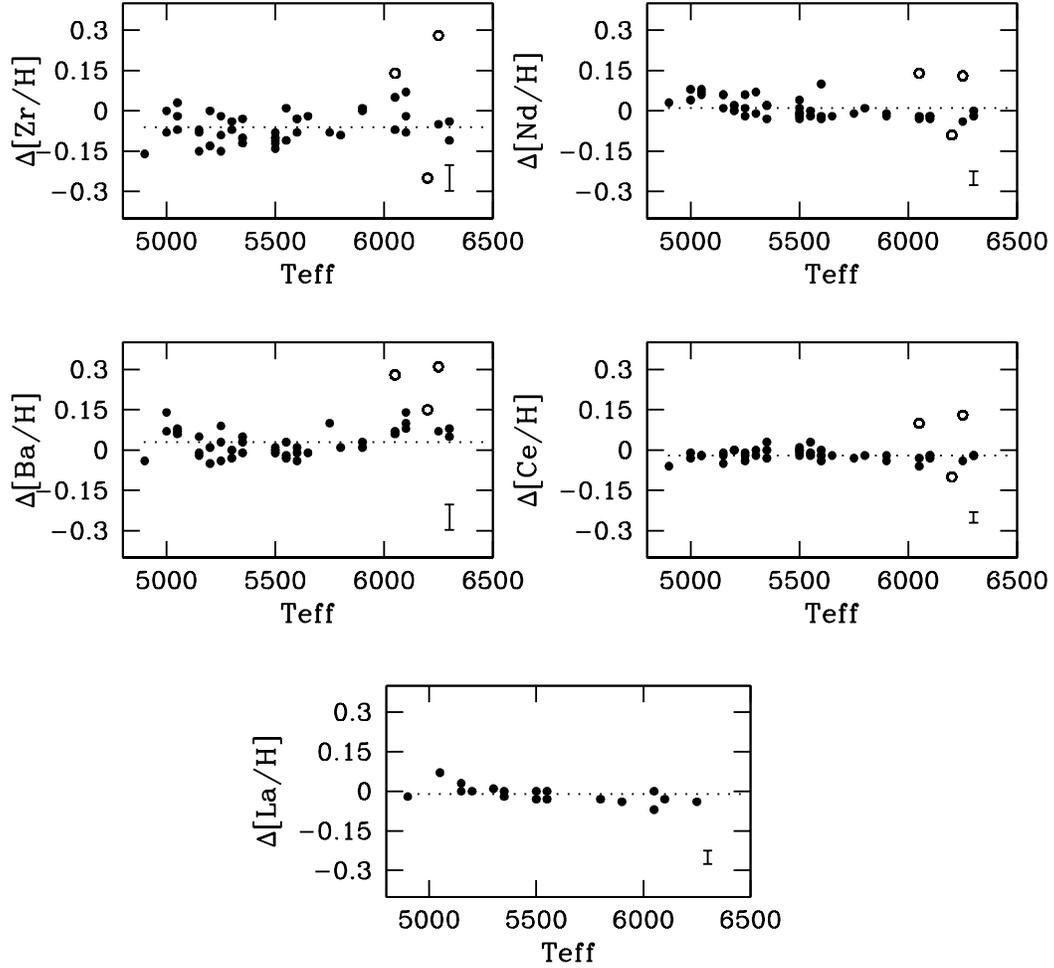}
\caption{\small Differential [X/H] vs. effective temperature. The differential
  comparision star is vB153 as discussed in the text. The open symbols
are stars believed to be non-members as discussed in Section
\ref{nm}. The dashed lines are mean abundance levels. The error bars shown is the typical abundance uncertainty for each element.}
\label{abplot}
\end{center}
\end{figure}

Our analysis produced absolute abundances, but to answer the
question of chemical homogeneity in the Hyades, we use differential
abundances. The final differential abundances $\Delta$[X/H] were derived by subtracting the  
absolute abundance of each individual line of the reference star vB153
from the same line in the sample stars, and then taking the mean of
the differences for each element. The star vB153 was chosen for
reference as it was the reference star used in the differential
abundance work by Pa03 hence enabling easy
comparisons. By using such a line-by-line differential technique we
also reduce the errors due to the uncertainty in the line data,
hence minimizing the star-to-star scatter. The differential abundances
of all elements are plotted in Figure \ref{abplot}. The absolute abundances
for the reference star vB153 is given in Table \ref{vb153}.

\subsection{Error Analysis}\label{errors}
The main sources of error in the present study are that of EW
measurements, continuum placement and stellar parameters. External
errors, such as uncertainties in the line data and model atmospheres are the
least sources of error since we are interested only in the differential
abundances. The number of lines used to calculate the final abundances also
contributed to the total uncertainty for each element. 

\smallskip

The error in EWs estimated by repeated measurements of each line, is between
0.5m\AA\ to 5m\AA\, depending 
on the strength of the lines, corresponding to abundance errors of 0.01 dex to
0.05 dex. The measurement errors for the synthesised abundances were derived by
changing the abundance until there is a clear visible deviation from the best
fit. The differential error in the stellar parameters were assumed to be
 $\delta T_{eff}$ = 50K, $\delta$log \emph{g} = 0.1 cm\,s$^{-2}$ and $\delta \xi$
= 0.1 \kms. Table \ref{parameters} shows the abundance dependence on the stellar
parameters. Typical values of the total estimated uncertainty in each element
is as follows; Zr = 0.048 dex, Ba = 0.047 dex, La = 0.025 dex, Ce = 0.025 dex, and Nd
= 0.026 dex.

\section{ABUNDANCE TRENDS}

The final results show uniform abundances although some elements show a small
residual trend with temperature. Slightly increasing slopes are seen for Zr,
while decreasing trends are seen for La and Nd. It is very likely
that these trends are a systematic effect resulting from inadequecies in the
model atmospheres, and does not reflect intrinsic abundance variations of the
Hyades cluster.  

\smallskip

\subsection{Stellar Parameters}

The abundances for the $\alpha$ and Fe-peak elements by Pa03
show no trends such as seen here for the neutron-capture elements (except for Ca
which has an increasing slope with temperature thought to be due to saturating
Ca lines in the cooler stars). Since we have used the same stellar parameters, it
is unlikely that errors in the parameters caused the presently observed
trends. Furthermore, any change in the model parameters will not correct all the
trends as different elements show trends in opposite directions. As a check,
each parameter ($T_{eff}$, log \emph{g}, and micro-turbulence) was varied in
an attempt to remove the trends. For effective temperature, the coolest stars
were made cooler and hottest stars made hotter by $\delta T_{eff}$ =
100K. Similarly the surface gravity was varied by $\delta$log \emph{g} = 0.1 cm\,s$^{-2}$ and micro-turbulence by $\delta \xi$
= 0.1 \kms. Such a change will \emph{flatten} the slopes for La and Nd, while the
same changes increase the slopes of Ba and Zr abundances.

\smallskip

\begin{figure}
\begin{center}
\includegraphics[scale=0.4, angle=0]{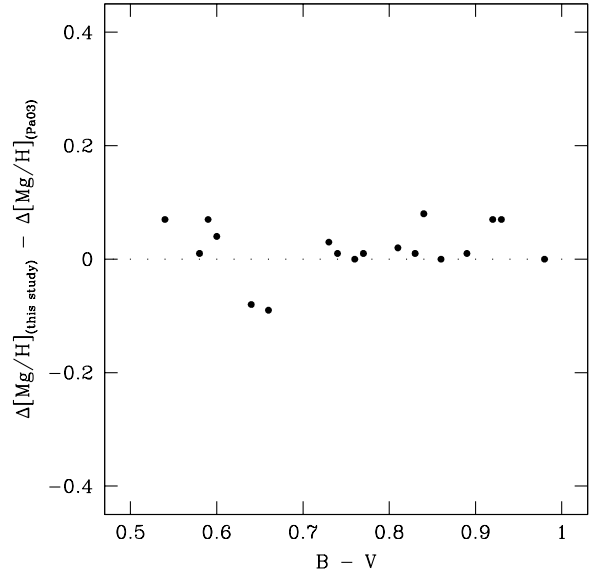}
\caption{Comparison of Mg abundances with Pa03 for a sub-sample of stars
covering the full temperature range. The mean of the difference $\Delta$[Mg/H]$_{\mbox{this study}}$ - $\Delta$[Mg/H]$_{\mbox{Pa03}}$
is 0.02 dex, with a standard deviation of 0.03 dex, but most importantly no slope is present.}\label{mgplot}
\end{center}
\end{figure}

Also to further confirm of our method of analysis, we re-analyzed the Mg
abundances for a sub-sample of stars covering the full temperature range, and
successfully recovered the results of Pa03, which does not show any trends. A
comparison plot is shown in Figure \ref{mgplot}. The mean of the difference
$\Delta$[Mg/H]$_{\mbox{this study}}$ - $\Delta$[Mg/H]$_{\mbox{Pa03}}$
= 0.02 dex, with a standard deviation of 0.03 dex.

\smallskip

\subsection{Continuum Placement}
Note that the scatter for all elements increase with lower temperature. As
the temperature decreases, the greater the blending and therefore the harder
it is for
continuum placement.  Since most lines lie in the blue part of the spectral
region which is highly crowded, determining the continuum level was
difficult. It is possible that as the lines begin to blend in the cooler
stars, the continuum is placed too low and a lower abundance is estimated for the
cooler stars in comparison to the stars at the hotter end. Such an effect would
explain the increasing trends observed for Zr, but we cannot explain the decreasing
slopes as due to an error in continuum placement.

\subsection{Hyperfine Structure}
Many of the heavy element lines are affected by hyperfine
structure. In most cases the effects are small enough
to be negligible, especially in very weak lines. However as lines saturate and
move off the linear part of the curve of growth, these effects become more
important. If an observed line is made up of many components, which are
greater than the intrinsic width of the line as calculated by standard means
of broadening (thermal, micro-turbulence, etc.) then neglecting the splitting
would result in over estimating the abundance value. The transition lines for
La, which has widely known hyperfine structure have been synthesised
accordingly. However a slight slope is still present in the La results after
accounting for the hyperfine structure.

\subsection{NLTE Effects}
Finally we suspect that NLTE effects may play a role in these abundance
trends. NLTE effects are specific to the atom and affect each element
differently rather than generally (cf. stellar parameters). Since we observe
slopes in opposite directions, it is very likely that we are seeing NLTE
effects. Unfortunately there have not been many NLTE calculations  
done for the heavier neutron-capture elements in order to determine
what the such effects would be. \citet{mashonkina99} find that for the
BaII line at 5853\AA\ the correction is less than 0.1 dex, however it does not
cover the full range of temperatures we are covering in this analysis. We are more interested to see if
NLTE effects come into play over a large range of temperatures in
order to explain the observed abundance trends. We are not aware of any
published work that indicate such NLTE effects. 

\smallskip

A study of NLTE effects for heavy elements is beyond the scope of this
paper, but in an attempt to quantify this possible NLTE correction we have fitted least
square regression lines to the observed trends. The fits are in the form of $ \Delta
\mbox{[X/H]} = \alpha + \beta\mbox{(5200K/$T_{eff}$)}$ where T$_{eff}$ has been
normalised by the T$_{eff}$ of the reference star vB153. The values of $\alpha$ 
and $\beta$ are given for each element in Table \ref{regpara}.  We look
forward to future NLTE calculations which may explain the presently 
observed abundance trends.

\section{DISCUSSION}
\subsection{Chemical Homogeneity}\label{chemicalhomo}
With regards to chemical homogeneity and establishing chemical
signatures for the Hyades, our results are positive. The tightly correlated
abundances are an indication of the level of homogeneity that exists within
the cluster. The observed scatter for the Hyades members
$\sigma_{obs}$, estimates of the intrinsic scatter $\sigma_{int}$, and the
corresponding Chi squared value, is given in Table
\ref{scatter}. $\sigma_{int}$ was estimated using the equation, $\sigma_{obs}^{2}
= \sigma_{int}^{2} + \sigma_{expected}^{2}$, and $\chi^{2}_{r} =
\sigma_{obs}^{2} / \sigma_{expected}^{2}$, where $\sigma_{expected}$ is the
estimated uncertainty as discussed in Section \ref{errors}. With the
estimated uncertainties being typically around $\sigma_{obs}$, then
$<\chi^{2}_{r}> \approx$ 1. This implies that the total star-to-star scatter
is within the measurable limit and that the true intrinsic scatter among the
Hyades stellar members are extremely low.  

\smallskip

This level of chemical uniformity is observational evidence for a chemically
well mixed gas cloud. It is possible that
these dwarfs formed later after the high mass stars have evolved to produce
several supernova, which enriched the gas as well as contributed to the mixing
of the gas cloud. Although the actual mechanism for the mixing is still
debatable, it is clear from our results that the gas cloud was uniformly mixed
by the time these dwarf stars formed.

\smallskip

Also, the detected levels of chemical homogeneity indicates that these stars
have not been \emph{polluted} either from local sources (such as stellar
winds, supernova explosions), or from swept up gas from the
ISM. \citet{quillen} derived an upper limit of 0.03 dex for the star-to-star
metallicity scatter using the Hyades H-R diagram, indicating that
pollution effects are not common and strong. Our estimated $\sigma_{int}$
for the heavier elements is within this upper limit.

\subsection{Stellar Membership}\label{nm}
Further reasons to interpret the results as a strong case for homogeneity and
positive signs of establishing a chemical signature is the deviation of a few
stars that are thought to be non-members found in the same direction
of the cluster. The two stars vB1 and vB2 are clearly over abundant by over
2$\sigma$ in all elements. Pa03 also found these 
stars to be significantly enriched. These stars were earlier considered as
Hyades members as they have consistent photometry, radial velocities, and
Hipparcos distances for membership, however \citet{debruijne01} find that
they are non-members based on proper motions and trigonometric parallax
measurements. The star HD14127 is under abundant in Zr, Ce, La, and Nd,
however it is slightly over abundant in Ba. Pa03 classes it as a non-member
due to its metallicity being -0.25 dex below the cluster mean, its
Hipparcos distance being too large, and it lies below the Hyades main
sequence. 

\subsection{Implications for Chemical Tagging}
Since most of the dynamical information of the Galactic disk has been subject
to evolutionary processes of dissipation and scattering, chemical
information locked within the stars provide the only true tracer
left of initial identity. Having established the Hyades open cluster members
to be homogeneous in the heavier neutron-capture elements along with the
$\alpha$ and Fe-peak elements (Pa03), and the fact that several non-members
deviate from the cluster mean, supports the case for using 
chemical abundances as tracers of cluster membership and establishing
chemical signatures unique only to the stellar members of 
the cluster. We are more confident now that chemical information
can be used to identify the larger star-forming aggregates in the early disk
by means of chemical tagging.

\smallskip

This paves the way for chemically tagging other member
stars which are at present no longer part of the bound Hyades cluster
system. A necessary test would be to study the stars of the
Hyades super-cluster, which is thought to be a moving stellar group
\citep[cf.][]{eggen70}. It will be very interesting indeed to check if the same
abundance patterns are seen among the stellar members of the super-cluster as
in the core of the Hyades open cluster we 
have presented here. Such a test will be the next step forward in truly
demonstrating the viability of chemical tagging for the future.

\acknowledgments

This research has made use of the Vienna Atomic Line Database (VALD), operated
at Vienna, Austria, and the Database on Rare Earths At Mons University
(D.R.E.A.M.), operated at Mons, Belgium.


\clearpage
\newpage

\clearpage
\newpage

\clearpage
\newpage

\clearpage
\newpage

\clearpage
\newpage

\begin{deluxetable}{lcccc}
\tabletypesize{\scriptsize}
\tablecaption{Line List\label{lines}}
\tablewidth{0pt}
\tablehead{
\colhead{Element} & \colhead{Wavelength} &
\colhead{$\chi$ (eV)} &
\colhead{log $gf$} &\colhead{Ref} }
\startdata
 \small ZrII    & 4050.33 &  0.710  &   -1.000 & 1\\
 \small BaII	& 4130.57 &  2.722  &    0.680 & 1\\
 \small BaII	& 4554.04 &  0.000  &    0.170 & 1\\
 \small BaII	& 4934.09 &  0.000  &   -0.150 & 1\\
 \small BaII	& 5853.69 &  0.604  &   -1.000 & 1\\
 \small BaII	& 6141.73 &  0.704  &   -0.076 & 1\\
 \small LaII    & 3988.52 &  0.400  &    0.210 & 2\\
 \small LaII    & 3995.75 &  0.170  &   -0.060 & 2\\ 
 \small LaII	& 4086.71 &  0.000  &   -0.070 & 2\\
 \small LaII	& 4662.50 &  0.173  &   -1.240 & 2\\
 \small CeII	& 4083.22 &  0.701  &    0.270 & 3\\
 \small CeII	& 4364.65 &  0.495  &   -0.230 & 3\\
 \small CeII	& 4562.36 &  0.478  &    0.230 & 3\\
 \small CeII	& 4628.16 &  0.516  &    0.200 & 3\\
 \small NdII	& 4012.24 &  0.630  &    0.810 & 4\\
 \small NdII	& 4012.69 &  0.000  &   -0.600 & 4\\
 \small NdII	& 4018.82 &  0.063  &   -0.850 & 4\\
 \small NdII	& 4068.89 &  0.000  &   -1.420 & 4\\
 \small NdII	& 4462.98 &  0.559   &   0.040 & 4\\
\enddata
\tablerefs{
(1) From Vienna Atomic Line Database (VALD) \citep{VALD1,VALD2,VALD3}.
(2) \citealt{lbs01}
(3) Palmeri et al. 2001 (http://www.umh.ac.be/$\tilde{}$astro/dream.shtml)
(4) \citealt{dlsc03} }
\end{deluxetable}

\clearpage
\begin{deluxetable}{lccccc}
\tabletypesize{\scriptsize}
\tablecaption{Elemental Abundances for vB153\label{vb153}
}
\tablewidth{0pt}
\tablehead{ \colhead{} &
\colhead{[Zr/H]} & \colhead{[Ba/H]} & \colhead{[La/H]} &
\colhead{[Ce/H]} & \colhead{[Nd/H]} }
\startdata
vB 153  &2.73  &2.60  & 1.19 &1.77 &1.45\\
\enddata
\end{deluxetable}

\begin{deluxetable}{lccccccccc}
\tabletypesize{\scriptsize}
\tablecaption{Abundance Dependencies on Model Parameters\label{parameters}
}
\tablewidth{0pt}
\tablehead{
\colhead{Star ID}& \colhead{Model Parameter}& \colhead{$\delta$[Zr/H]} & 
\colhead{$\delta$[Ba/H]} & \colhead{$\delta$[La/H]} & 
\colhead{$\delta$[Ce/H]} & \colhead{$\delta$[Nd/H]} } 

\startdata

vB 25 &T$_{\rm eff}\pm$50 & $\pm$0.01 & $\pm$0.02 & $\pm$0.01 & $\pm$0.01 & $\pm$0.02\\
($T_{\rm eff}$ = 4900)&log $g$$\pm$0.1 & $\pm$0.03 & 0.00 & $\pm$0.01 & 0.00 & 0.00\\
&$\xi\pm$0.1 & $\mp$0.02 & $\mp$0.03 & $\pm$0.02 & 0.00 & $\pm$0.02 \\ \hline

vB 105 &T$_{\rm eff}\pm$50 & $\pm$0.01 & $\pm$0.02 & $\pm$0.01 & $\pm$0.01 & 0.00 \\
($T_{\rm eff}$ = 6100)&log $g$$\pm$0.1& $\pm$0.04 & $\pm$0.01 & 0.00 & $\pm$0.01 & $\pm$0.02 \\
&$\xi\pm$0.1 &$\mp$0.02 & $\mp$0.03 & $\pm$0.01 & 0.00 & 0.00 \\
\enddata
\end{deluxetable}

\begin{deluxetable}{lcc}
\tabletypesize{\scriptsize}
\tablecaption{Regression Line Parameters\label{regpara}}
\tablewidth{0pt}
\tablehead{
\colhead{Element} & \colhead{$\alpha$} &
\colhead{$\beta$} }
\startdata
Zr II & 2.88 &  -0.22\\
Ba II & 2.81 &  -0.19\\
La II & 0.90 &   0.29\\
Ce II & 1.72 &   0.04\\
Nd II & 1.11 &   0.37\\
\enddata
\end{deluxetable}

\begin{deluxetable}{lcccc}
\tabletypesize{\scriptsize}
\tablecaption{Abundance Scatter\label{scatter}}
\tablewidth{0pt}
\tablehead{
\colhead{Element} & \colhead{$\sigma_{obs}$} &
\colhead{$\sigma_{int}$}  & \colhead{$\chi^{2}_{r}$} }
\startdata
Zr & 0.055 & 0.026 &1.30 \\
Ba & 0.049 & 0.013 &1.08 \\
La & 0.025 & 0.000 &1.00 \\
Ce & 0.025 & 0.000 &1.00 \\
Nd & 0.032 & 0.018 &1.51 \\
\enddata
\end{deluxetable}

\end{document}